\documentclass[10pt,twocolumn,letterpaper]{article}
\usepackage{graphicx}
\usepackage[dvipdfmx]{color}

\usepackage{multirow}
\usepackage{color}
\usepackage{colortbl}
\usepackage{lineno}
\usepackage{setspace}
\usepackage{amsmath}
\usepackage{amssymb}

\usepackage[pagebackref=true,breaklinks=true,letterpaper=true,colorlinks,citecolor=blue,linkcolor=blue,bookmarks=false]{hyperref}

\begin{document}
\title{Image Super-Resolution using Explicit Perceptual Loss}

\author{
  Tomoki Yoshida, 
  Kazutoshi Akita,
  Muhammad Haris,
  Norimichi Ukita\\
  Toyota Technological Institute, Japan
}

\date{}

\maketitle

\section*{\centering Abstract}
\textit{
This paper proposes an explicit way to optimize the super-resolution network for generating visually pleasing images.
The previous approaches use several loss functions which is hard to interpret and has the implicit relationships to improve the perceptual score.
We show how to exploit the machine learning based model which is directly trained to provide the perceptual score on generated images.
It is believed that these models can be used to optimizes the super-resolution network which is easier to interpret.
We further analyze the characteristic of the existing loss and our proposed explicit perceptual loss for better interpretation.
The experimental results show the explicit approach has a higher perceptual score than other approaches.
Finally, we demonstrate the relation of explicit perceptual loss and visually pleasing images using subjective evaluation.
}

\section{Introduction}
\label{section:introduction}

Single-image Super-resolution (SR)~\cite{Irani:1991:IRI:108693.108696}
enlarges a low-resolution image (LR) so that its image quality is maintained.
As with many computer vision technologies, SR has been improved
significantly with convolutional neural networks, CNNs (e.g.,
DBPN~\cite{DBLP:conf/cvpr/HarisSU18,DBLP:journals/corr/abs-1904-05677},
WDST~\cite{DBLP:conf/iccv/0002YXD19},
and
SRFlow~\cite{eccv2020sisr}).
Its performance is improved every year as demonstrated in public
challenges~\cite{DBLP:conf/cvpr/TimofteGWG18,DBLP:conf/iccvw/GuLZXYZYSTDLDLG19,DBLP:conf/cvpr/ZhangGTSDZYGJYK20}.
In the common SR methods using CNNs as well as those without CNNs, SR models are
trained so that the mean square error (MSE) is minimized. The MSE is
computed from the difference between a reconstructed SR image and its
high-resolution (ground-truth) image.

However, it is revealed that the MSE minimization leads to
perceptually-discomfortable SR images
\cite{DBLP:conf/iccv/SajjadiSH17, DBLP:conf/cvpr/LedigTHCCAATTWS17}.
In these works, perceptually-comfortable images are
reconstructed additional loss functions such as perceptual loss~\cite{johnson2016perceptual}, adversarial loss~\cite{DBLP:conf/cvpr/LedigTHCCAATTWS17}, and style
loss~\cite{DBLP:conf/iccv/SajjadiSH17}.
In~\cite{DBLP:conf/cvpr/BlauM18}, it is demonstrated that there exists
a trade-off between the image-distortion quality evaluated by the MSE
and the perceptual quality.
%

In these approaches~\cite{DBLP:conf/iccv/SajjadiSH17,
  DBLP:conf/cvpr/LedigTHCCAATTWS17}, the perceptual quality is
improved implicitly by several loss functions whose relationship with
the perceptual score is difficult to be interpreted. The difficulty in
this interpretation is increased due to deep networks in the SR
methods~\cite{DBLP:conf/iccv/SajjadiSH17,
  DBLP:conf/cvpr/LedigTHCCAATTWS17} described above.

On the other hand, we can explicitly improve the perceptual quality of
machine learning (ML) based SR models by simpler ways.
The most straightforward way may be to manually provide subjective
perceptual scores to all possible SR images that are generated and
evaluated during the supervised training stage. Unfortunately, that
is impossible in reality.
Such explicit perceptual scores, however, can be predicted by
perceptual-quality-aware features~\cite{DBLP:journals/spl/MittalSB13}
and ML models that are trained directly by the subjective perceptual
scores~\cite{DBLP:journals/cviu/MaYY017}.
These features and models can be utilized for perceptual loss
functions that explicitly improve the perceptual quality.

\begin{figure*}[!t]
  \begin{center}
    \begin{tabular}[c]{ccccc}
      \includegraphics[width=.17\textwidth]{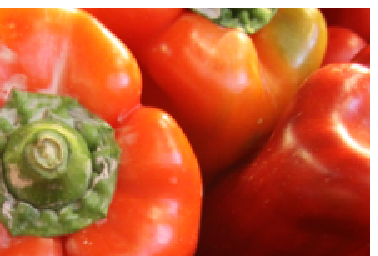}
      &
        \includegraphics[width=.17\textwidth]{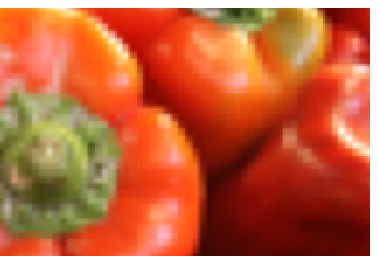}
      &
        \includegraphics[width=.17\textwidth]{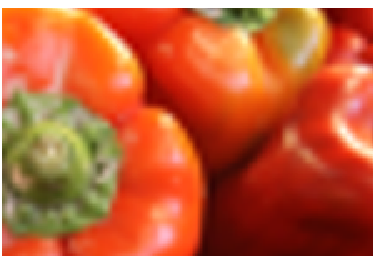}
      &
        \includegraphics[width=.17\textwidth]{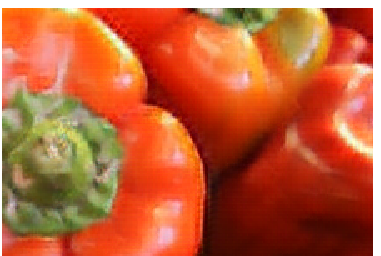}
      &
        \includegraphics[width=.17\textwidth]{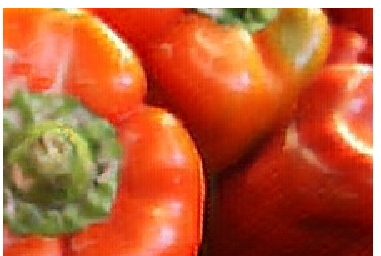}\vspace{-0.2em}\\
        
      {\small (a) GT}
      &{\small (b) LR}
      &{\small (c) Bicubic}
      &{\small (d) Implicit}
      &{\small (e) Explicit}\vspace{0.2em}\\
      
      &&Perc.: 7.36 &Perc.: 3.90 &Perc.: 3.77
    \end{tabular}\vspace{1em}
    \caption{Effect of explicit perceptual loss functions on SR. Noted: Perc. is \texttt{Perceptual score} where a lower score indicate better result.}
    \label{fig:effect}\vspace{-1em}
  \end{center}
\end{figure*}

In this paper, we evaluate the effectiveness of the aforementioned loss
functions for implicit and explicit improvement of the perceptual SR
quality as briefly shown in Fig.~\ref{fig:effect}. 
The explicit perceptual loss is able to improve the perceptual score compare with other approaches.

\section{Related Work}
\label{section:related}

Image restoration and enhancement including image SR require
appropriate quality assessment metrics for evaluation. Such metrics
are important also for training objectives, if the quality assessment
is given by ML with training data.
PSNR and SSIM~\cite{DBLP:journals/tip/WangBSS04} are widely used as
such metrics, focusing on comparing a reconstructed image with its
ground truth image.
There exist methods for quality assessment that do not require a
reference ground truth image~\cite{DBLP:conf/icip/Luo04,
  DBLP:conf/cvpr/TangJK14, DBLP:journals/spl/MittalSB13}, including
some that use deep neural networks to learn the metrics
\cite{DBLP:conf/cvpr/KangYLD14, DBLP:journals/tip/MaLZDWZ18}.

Several quality assessment metrics~\cite{DBLP:conf/icip/ReibmanBG06,
  DBLP:conf/icip/YeganehRW12, DBLP:conf/icip/FangLZLG16} have been
evaluated specifically for SR, including no-reference metrics (a.k.a
blind metrics)~\cite{DBLP:journals/cviu/MaYY017}.

For the goal of this paper (i.e., SR model learning), no-reference
metrics are required because any SR images reconstructed by any
intermediate learning results of model parameters must be evaluated.
Among the aforementioned no-reference metrics, NIQE
\cite{DBLP:journals/spl/MittalSB13} and Ma's algorithm
\cite{DBLP:journals/cviu/MaYY017} are regarded as the representatives
of explicit perceptual metrics
based on hand-crafted features and ML, respectively, and
utilized in the perceptual-aware SR competition
\cite{DBLP:journals/corr/abs-1809-07517}.
These metrics~\cite{DBLP:journals/spl/MittalSB13,
  DBLP:journals/cviu/MaYY017} are explained briefly as our explicit
perceptual loss functions at the beginning of the next section.

\section{SR Image Reconstruction based on Explicit Perceptual Loss}
\label{section:method}

\subsection{No-reference Perceptual Quality Assessment Metrics for Explicit Perceptual Metrics}
\label{subsection:metrics}

\paragraph{NIQE}

NIQE~\cite{DBLP:journals/spl/MittalSB13}, which is one of the explicit
perceptual loss used in our experiments, uses a collection of
quality-aware statistical features based on a simple and successful
space domain natural scene statistic (NSS) model. The NSS model is
appropriate for representing the explicit perceptual loss because this
model describes the statistics to which the visual apparatus has
adapted in natural images.

For NIQE, the NSS feature~\cite{bib:NSS},
$\frac{I (i, j) - \mu (i, j)}{\sigma (i, j) + 1},$ is computed in all
pixels of an image,
where $I(i, j)$, $\mu (i, j)$, and $\sigma (i, j)$ denote a pixel
value in $(i, j)$ and the mean and variance of pixels around $(i, j)$,
respectively. In NIQE, $7 \times 7$ pixels are used for computing
$\mu(i, j)$ and $\sigma(i, j)$.
Only patches where the sum of $\sigma(i ,j)$ is above a predefined
threshold are used in the following process.

In each patch, parameters in the following two Gaussian distributions
of pixel values $x$ are computed:
\begin{itemize}
\item {\bf Generalized Gaussian distribution (GGD):}
\begin{equation}
f_{g} \left( x; \alpha, \beta \right) = \frac{\alpha}{2 \beta
  \Gamma(1/\alpha)} \exp \left( - \left( \frac{|x|}{\beta}
  \right)^{\alpha} \right),
\label{eq:ggd}
\end{equation}
where $\gamma$ is the gamma function. $\alpha$ and $\beta$ are
estimated as the parameters.
\item {\bf Asymmetric generalized Gaussian distribution (AGGD):} 
\footnotesize
\begin{equation}
f_{a} \left( x; \gamma, \beta_{l}, \beta_{r} \right) =
\frac{\gamma}{(\beta_{l} + \beta_{r})
  \Gamma(1/\gamma)} \exp \left( - \left( \frac{|x|}{\beta'} \right)^{\gamma} \right),
\label{eq:aggd}
\end{equation}
\normalsize
where $\beta'$ is $\beta_{l}$ (when $x \leq 0$) or $\beta_{r}$ (when
$x \geq 0$). $\gamma$, $\beta_{l}$, and $\beta_{r}$ are estimated as
the parameters. The mean of the distribution, $\eta$, is also
parameterized:
\begin{equation}
\eta = \left( \beta_{l} - \beta_{r} \right) \frac{\gamma(\frac{2}{\gamma})}{\Gamma(\frac{1}{\gamma})}
\label{eq:eta}
\end{equation}
\end{itemize}

$\gamma$, $\beta_{l}$, $\beta_{r}$, and $\eta$ in Eqs. (\ref{eq:aggd})
and (\ref{eq:eta}) are estimated along the four orientations and used
in conjunction with $\alpha$ and $\beta$ in Eq. (\ref{eq:ggd}). In
total, 18 parameters are given in each patch.

The multivariate distribution of the estimated 18 features is
represented by the multivariate Gaussian (MVG) model.
With the mean and variance of the MVG, the quality of the test image
is evaluated by the following distance between the MVG models of
natural images and the test image: \footnotesize
\begin{equation}
D(\nu_{n}, \nu_{t}, \Sigma_{n}, \Sigma_{t}) = \sqrt{\left( (\nu_{n} - \nu_{t})^{T} \left( \frac{\Sigma_{1} + \Sigma_{2}}{2} \right)^{-1} (\nu_{n} - \nu_{t}) \right)},
\label{eq:D}
\end{equation}
\normalsize
where $\nu_{n}, \nu_{t}$ and $\Sigma_{n}, \Sigma_{t}$ are the mean
vectors and covariance matrices of the MVG models of the natural
images and the test image, respectively. In NIQE, the natural images
were selected from Flickr and the Berkeley image segmentation database.

\paragraph{Ma's Algorithm}

Ma's algorithm~\cite{DBLP:journals/cviu/MaYY017} evaluates the
perceptual quality of an image based on three types of low-level
statistical features in both spatial and frequency domains, namely the
local frequency features computed by the discrete cosine transform,
the global frequency features based on represented by the wavelet
coefficients, and the spatial discontinuity feature based on the
singular values.
%
For each of the three statistical features, a random forest regressor
is trained to predict the human-subjective score of each training
image.

While the original algorithm~\cite{DBLP:journals/cviu/MaYY017}
predicts the perceptual score by a weighted-linear regression using
the aforementioned three features, the computational cost of the local
frequency feature is too huge; 20 times slower than other two
features. This huge computational cost is inappropriate for a loss
function in ML.
Furthermore, the contribution of the global frequency feature is
smaller compared with the spatial discontinuity feature, as
demonstrated in~\cite{DBLP:journals/cviu/MaYY017}.
In our experiments, therefore, the spatial discontinuity feature is
evaluated as one of the explicit perceptual loss in order to avoid the
combinatorial explosion in evaluation with NIQE and three implicit
perceptual loss functions~\cite{DBLP:conf/iccv/SajjadiSH17,
  DBLP:conf/cvpr/LedigTHCCAATTWS17}; while we use the five loss
functions resulting in $2^{5}=32$ combinations in our experiments
(Section~\ref{subssec:multi_loss}), $2^{7}=128$ tests are required if
all of the seven loss functions including the local and global
frequency features are evaluated.

Singular values of images with smooth contents become zero more
rapidly than for those with sharp contents, as validated in~\cite{DBLP:journals/cviu/MaYY017}.
This property suggests us to use the singular values for evaluating
the spatial discontinuity.
In the Ma's spatial discontinuity (MSD) metric, the singular values
are computed from a set of patches that are extracted from the image
with no overlaps. The concatenation of the singular values is used as
a MSD feature vector and fed into the random forest regressor.

\subsection{SR Loss Functions using Explicit Perceptual Metrics}
\label{subsection:method}

\begin{figure}[t]
  \begin{center}
    \includegraphics[width=\columnwidth]{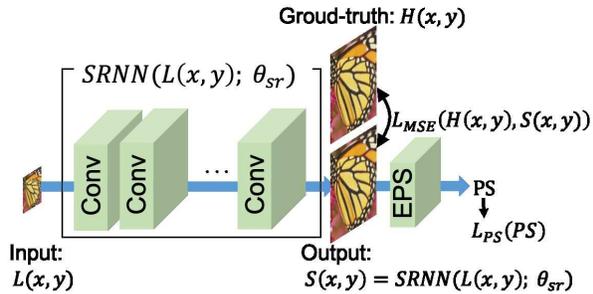}
    \vspace*{1mm}
    \caption{Deep SR network with the MSE loss ($L_{MSE}$) and our
      explicit perceptual loss ($L_{PS}$). ``EPS'' in the figure
      outputs the explicit perceptual score of each reconstructed SR
      image, $S(x,y)$. $\theta_{sr}$ denotes the parameters of the
      network.}
    \label{fig:network}
  \end{center}
  \vspace*{-5mm}
\end{figure}

In this section, we propose how to use the two explicit perceptual
metrics, NIQE and MSD, as loss functions that are applicable to deep
SR networks.
A simple illustration of the SR network is shown in
Figure~\ref{fig:network}. Basic SR networks such as SRCNN and DBPN
employ only the MSE loss, which is
$L_{MSE} = \frac{1}{XY} \sum^{X,Y}_{x,y} \left( S(x,y) - H(x,y)
\right)^{2}$
in this figure where $S(x,y)$ and $H(x,y)$ denote the SR and
ground-truth high-resolution (HR) images, each of whose size is
$(X, Y)$, respectively.

Since NIQE outputs a scalar score (i.e.,
$D(\nu_{n}, \nu_{t}, \Sigma_{n}, \Sigma_{t})$ in (\ref{eq:D})) only
based on statistic calculation,
$D(\nu_{n}, \nu_{t}, \Sigma_{n}, \Sigma_{t})$ can be directly used as
a loss function for deep SR networks.
In this case, ``EPS'' in Figure~\ref{fig:network} consists of
Equations (\ref{eq:ggd}), (\ref{eq:aggd}), (\ref{eq:eta}), and
(\ref{eq:D}).

In MSD, on the other hand, a ML based regressor is
employed at the final stage.
Any type of regressors can be employed, including deep networks,
random forests, support vector machines, and so on.  For example, the
Ma's algorithm uses random forest regressors for all metrics.
Depending on the type of the regressor, we propose the following two
options for using MSD as a loss function:
\begin{description}
\item[Deep networks:] As with NIQE, the MSD regressor using a deep
  network that outputs a scalar MSD score can be straightforwardly
  combined with deep SR networks.
  In this case, ``EPS'' in Figure~\ref{fig:network} indicates the MSD
  regressor using the deep network.
\item[Other ML algorithms:] We have difficulty in containing other ML
  algorithms into a deep network both for efficient training and
  inference. Our proposed method resolves this problem by employing a
  MSD feature vector instead of a MSD score. The MSD feature vector is
  computed from any image. In the training stage, we compute the MSE
  loss between the MSD feature vectors of each reconstructed image and
  its original HR image in ``EPS'' in Figure~\ref{fig:network}.
   This loss is used for evaluating the perceptual quality of each
  reconstructed SR image While the MSD feature vector is computed from
  every reconstructed SR image, the one of the HR image is computed
  only once at the beginning of the training stage.
\end{description}

Finally, the loss function with NIQE and MSD is defined as follows:
\begin{equation}
  L_{PS} = \epsilon \left( D(\nu_{n}, \nu_{t}, \Sigma_{n},
    \Sigma_{t}) \right)^{2}
  + \zeta L_{Ma},
  \label{eq:loss}
\end{equation}
where $\epsilon$ and $\zeta$ denote the weights of NIQE and MSD,
respectively, and $L_{Ma}$ is either of the following ones:
\begin{description}
\item[Deep networks:]
\begin{equation}
L_{Ma} = PSNN(M^{S}),
\label{eq:dn}
\end{equation}
where $PSNN(M^{S})$ denotes the deep network that regresses the
perceptual score from $M^{S}$, which is the MSD feature vector of the
reconstructed SR. $PSNN(M^{S})$ is trained so that a lower score means
the better perceptual quality.
\item[Other ML algorithms:]
\begin{equation}
L_{Ma} = \sum_{i} \left( M^{S}_{i} - M^{H}_{i} \right)^{2},
\label{eq:ml}
\end{equation}
where $M^{H}$ denotes the MSD feature vector of the original HR image.
\end{description}
These two options, Eqs (\ref{eq:dn}) and (\ref{eq:ml}), are evaluated
in Section~\ref{subssec:explicit_loss}.

\section{Experimental Results}
\label{section:experiments}

\subsection{Implementation and training details} 
\begin{figure}
  \begin{center}
  \includegraphics[width=.5\textwidth]{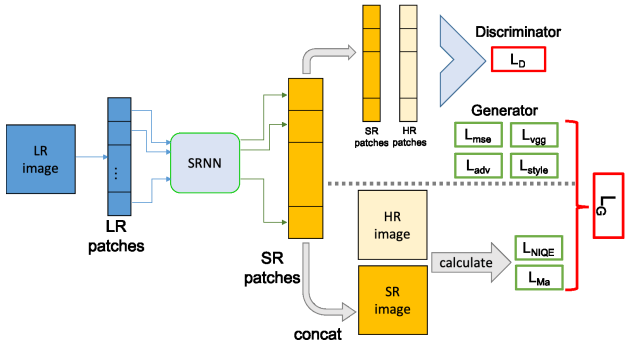}
  \end{center}\vspace{-1em}
  \caption{The overview of training mechanism.}
  \label{fig:train}
\end{figure}

The networks consist of two blocks: generator ($G$) and discriminator ($D$)~\cite{goodfellow2014generative}.
$G$ network generates (SR) images and $D$ network differentiates between real (HR) and fake (SR) images. 
DBPN~\cite{DBLP:conf/cvpr/HarisSU18}, the winner of SR competition held in
2018~\cite{DBLP:conf/cvpr/TimofteGWG18, pirm2018}, was used as $G$ network.
Meanwhile, $D$ network consists of five hidden layers with batch norm and the last layer is fully connected layer.
The training mechanism is illustrated in Fig.~\ref{fig:train}. 

We trained the networks using images from DIV2K \cite{Agustsson_2017_CVPR_Workshops} with online augmentation (scaling, rotating, flipping, and random cropping). 
To produce LR images, we downscale the HR images on particular scaling factors using Bicubic. 
We use the validation set from PIRM2018~\cite{pirm2018} as the test set which consists of 100 images.

The experiment focuses on 4$\times$ scaling factor.
We use DBPN-S which is the variant of DBPN with shallower depth.
On DBPN-S, we use $8 \times 8$ kernel with stride = 4 and pad by 2 pixels with T = 2.
All convolutional and transposed convolutional layers are followed by parametric rectified linear units (PReLUs), except the final reconstruction layer.
We initialize the weights based on~\cite{he2015delving}. 

We use batch size of 4 with size $576 \times 576$ for HR image, while LR image size is 144 $\times$ 144.
We intentionally use big size patches assuming explicit perceptual loss works better on bigger patch than on the smaller patches as shown in Fig.~\ref{fig:train}. 
The learning rate is initialized to $1e-4$ for all layers and decrease by a factor of 10 for every 50 epochs for total 100 epochs. 
We used Adam with momentum to $0.9$. 
All experiments were conducted using PyTorch 0.4.1 and Python 3.5 on NVIDIA TITAN X GPUs.

To evaluate the performance, we use \texttt{Perceptual score} proposed in PIRM2018~\cite{pirm2018}, the perceptual super-resolution challenge. It is divided into three categories defined by thresholds on the RMSE. The three region are defined by Region 1: RMSE $\le 11.5$ , Region 2: $11.5 <$ RMSE $\le 12.5$, and Region 3: $12.5 <$ RMSE $\le 16$. The \texttt{Perceptual score} is computed by combining the quality measures of Ma~\cite{DBLP:journals/cviu/MaYY017} and NIQE~\cite{DBLP:journals/spl/MittalSB13} as below. A lower score means the better perceptual quality.
\begin{equation}
\texttt{Perceptual score} = 1/2((10-Ma)+NIQE)
\end{equation}

\subsection{Combination on multiple loss functions}
\label{subssec:multi_loss}
Here, we evaluate the combination of five losses function on G network to show the characteristic of each loss, producing 32 combinations.
On G network, we implement six losses (MSE, VGG, Style, Adversarial loss, NIQE, and Ma) which explained as below. 
\begin{equation}\begin{split}
L_{G}  &= 10*L_{mse} + w_1*L_{vgg} + w_2*L_{adv} + \\
&w_3*L_{style} + w_4*L_{NIQE} + w_5*L_{Ma} 
\end{split}
\end{equation}

\begin{description}
\item[(a) $L_{mse}$] is pixel-wise loss $L_{mse} = || I^h - I^{sr} ||^2_2$. 
\item[(b) $L_{vgg}$] is calculated in the feature space using pretrained VGG19~\cite{simonyan2014very} on multiple layers. This loss was originally proposed by~\cite{johnson2016perceptual, dosovitskiy2016generating}. Both $I^h$ and $I^{sr}$ are first mapped into a feature space by differentiable functions $f_{i}$ from VGG multiple max-pool layers ${(i = {2, 3, 4, 5})}$ then sum up each layer distances. $L_{vgg} = \sum\limits_{i={2}}^5 || f_i(I^h) - f_i(I^{sr}) ||^2_2$.
\item[(c) $L_{adv}$] $= -\texttt{log} (D(G(I^{l})))$~\cite{goodfellow2014generative}
\item[(d) $L_{style}$] is used to generate high quality textures~\cite{gatys2016image}. 
\item[(e) $L_{NIQE}$] (Eqs.~\ref{eq:D})
\item[(f) $L_{Ma}$] (Eqs.~\ref{eq:loss})
\end{description}

For $D$ network~\cite{goodfellow2014generative}, it is optimized by 
\begin{equation}\begin{split}
L_D = -\texttt{log} (D(I^{h})) - \texttt{log}(1-D(G(I^{l})))
\end{split}
\end{equation}

Table~\ref{tab:multi_loss} shows the results on 32 combinations. 
It is hard to interpret the behavior of loss function. 
However, some results can be highlighted to generally understand the characteristic of each loss, especially between implicit and explicit perceptual loss. 
Noted that the weight for each loss is chosen based on preliminary experiments.

Among a single loss function (no.~1~-~6), it is shown that $L_{NIQE}$ provides the best results on Region 2.
Further improvement can be achieved when we combined explicit perceptual loss with $L_{adv}$ as shown in no.~12.
However, we start observing diminishing returns on $L_{NIQE}$ even combined with other implicit perceptual loss.
Meanwhile, $L_{Ma}$ shows good performance only if combined with $L_{vgg}$ and $L_{adv}$.

The best combination is shown on no.~19 which use four loss functions: $L_{mse}$, $L_{vgg}$,$L_{adv}$, and $L_{Ma}$.
It is also interesting to see that the second best result is no.~17 which also use four loss functions but replacing $L_{Ma}$ with $L_{style}$.
Therefore, we can conclude that $L_{Ma}$ is able to replace $L_{style}$ with a marginal improvement.

The best result on two explicit perceptual loss is shown by no.~31. 
However, it is important to note that $L_{adv}$ is crucial to improve the performance of this combination. 
We can clearly see it by comparing no.~31 and 30 where no.~30's performance is much worse by only eliminating $L_{adv}$.

\begin{table}[t!]
\scriptsize
\caption{The comparison of six losses on 32 combinations.}
  \begin{center}
\scalebox{0.95}{
\begin{tabular}{|*1c|*1c|*1c|*1c|*1c|*1c|*1c|*1c|*1c|}
\hline
No. & $w_1$ & $w_2$ & $w_3$ & $w_4$ & $w_5$ & \texttt{Perc.} & RMSE & Region \\     
\hline\hline
1&0&0&0&0&0&5.692&11.86&2\\\hline
2&0.1&0&0&0&0&5.654&11.82&2\\\hline
3&0&0.1&0&0&0&2.540&14.12&3\\\hline
4&0&0&10&0&0&5.699&11.76&2\\\hline
5&0&0&0&0.01&0&5.397&11.86&2\\\hline
6&0&0&0&0&0.001&5.666&11.89&2\\\hline
7&0.1&0.1&0&0&0&2.751&13.86&3\\\hline
8&0.1&0&10&0&0&5.784&12.1&2\\\hline
9&0.1&0&0&0.01&0&5.587&12.13&2\\\hline
10&0.1&0&0&0&0.001&5.713&11.81&2\\\hline
11&0&0.1&10&0&0&2.580&13.9&3\\\hline
12&0&0.1&0&0.01&0&2.575&13.73&3\\\hline
13&0&0.1&0&0&0.001&5.745&11.79&2\\\hline
14&0&0&10&0.01&0&5.557&11.98&2\\\hline
15&0&0&10&0&0.001&5.685&11.89&2\\\hline
16&0&0&0&0.01&0.001&5.506&11.94&2\\\hline
17&0.1&0.1&10&0&0&2.479&13.86&3\\\hline
18&0.1&0.1&0&0.01&0&2.562&14.44&3\\\hline
19&0.1&0.1&0&0&0.001&2.471&14.07&3\\\hline
20&0.1&0&10&0.01&0&5.733&11.82&2\\\hline
21&0.1&0&10&0&0.001&5.657&11.77&2\\\hline
22&0.1&0&0&0.01&0.001&5.533&11.85&2\\\hline
23&0&0.1&10&0.01&0&2.580&13.98&3\\\hline
24&0&0&10&0.01&0.001&5.459&11.85&2\\\hline
25&0&0.1&0&0.01&0.001&2.626&14.27&3\\\hline
26&0&0.1&10&0&0.001&3.089&13.80&3\\\hline
27&0.1&0.1&10&0.01&0&2.724&13.78&3\\\hline
28&0.1&0.1&10&0&0.001&2.549&13.78&3\\\hline
29&0.1&0.1&0&0.01&0.001&2.507&13.84&3\\\hline
30&0.1&0&10&0.01&0.001&5.614&11.86&2\\\hline
31&0&0.1&10&0.01&0.001&2.497&13.81&3\\\hline
32&0.1&0.1&10&0.01&0.001&2.537&14.03&3\\
\hline
\end{tabular}}
\label{tab:multi_loss}
  \end{center}
\end{table}

\subsection{Different type of regressor on explicit perceptual loss}
\label{subssec:explicit_loss}
We conduct experiment to evaluate two types of regressor on explicit perceptual loss.
Here, the network is only optimized by one loss, either NN~(\ref{eq:dn}) or other ML~(\ref{eq:ml}).
The results are shown in Table~\ref{tab:explicit_loss}.
(\ref{eq:dn}) approach has a marginal decline compare with (\ref{eq:ml}).
It can be assumed that (\ref{eq:ml}) performed better than (\ref{eq:dn}).
Furthermore, (\ref{eq:ml}) provide low computation and less hyperparameter are needed to ease the optimization process.

\begin{table}[t!]
\small
\caption{The comparison of different type of the regressor for explicit perceptual loss.}
\begin{center}
\begin{tabular}{*1c|*1c|*1c}
\hline
Method & $\texttt{Perceptual Score}$ & RMSE\\     
\hline
NN~(\ref{eq:dn}) & 5.729 & 11.83\\
other ML~(\ref{eq:ml}) & 5.666 & 11.90\\
\hline
\end{tabular}
\label{tab:explicit_loss}
\end{center}
\end{table}

\subsection{Subjective evaluation}
We performed a subjective test to quantify the performance of different combination of loss function. Specifically, we asked 30 raters to assign a score from 1 (bad quality) to 10 (best quality) to each image.
The raters rated 7 combinations of loss function: $L_{mse}$, $L_{mse}+L_{vgg}$, $L_{mse}+L_{adv}$, $L_{mse}+L_{style}$, $L_{mse}+L_{NIQE}$, $L_{mse}+L_{Ma}$, $L_{mse}+L_{vgg}+L_{adv}+L_{style}+L_{NIQE}+L_{Ma}$. In total, each raters rated 700 instances (7 combinations of 100 images).

\begin{figure}
  \begin{center}
  \includegraphics[width=.5\textwidth]{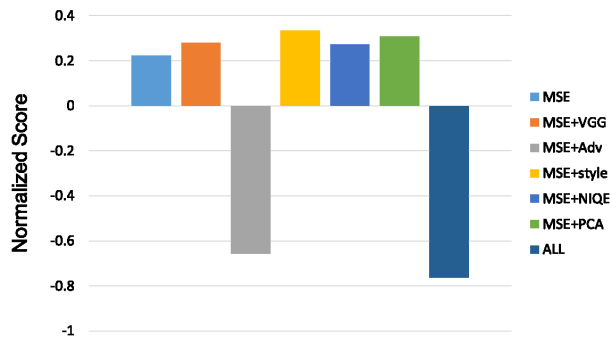}
  \end{center}\vspace{-1em}
  \caption{The result of subjective test.}
  \label{fig:subjective}
\end{figure}

The result of subjective evaluation is shown in Fig.~\ref{fig:subjective}.
Most of the raters give a lower score for ``MSE+Adv'' and ``All'', while there are slight differences between other methods. 
The best subjective score is achieved by ``MSE+Style''.
On Section~\ref{subssec:multi_loss}, it shows that explicit perceptual loss is able to improve the \texttt{perceptual score}.
However, the subjective evaluation shows that better \texttt{perceptual score} does not give better visualization on human perception.

\begin{figure}[!t]
  \begin{center}
    \begin{tabular}[c]{cc}
      \includegraphics[width=.17\textwidth]{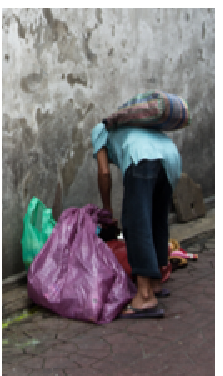}
      &
        \includegraphics[width=.17\textwidth]{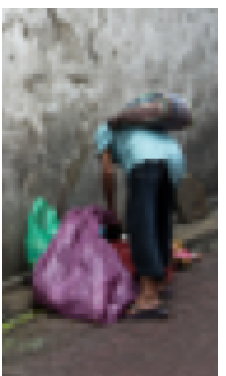}\\
              {\small (a) GT}
      &{\small (b) LR}\\
      
        \includegraphics[width=.17\textwidth]{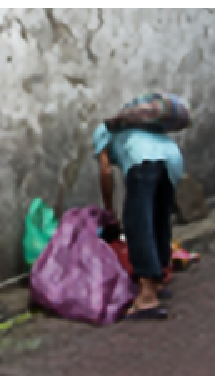}
      &
        \includegraphics[width=.17\textwidth]{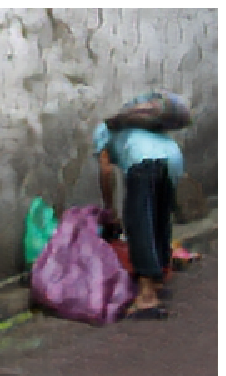}\\
        
      {\small (c) MSE}
      &{\small (d) Implicit}\vspace{0.2em}\\
      
    \end{tabular}\vspace{1.5em}
    \caption{The comparison of different loss function. The implicit perceptual loss tends to create high-frequency artifacts which can be considered as noise by human perception.}
    \label{fig:subject}
  \end{center}
\end{figure}

This result produces at least two observations.
From this evaluation, it shows there is no strong correlation between the existing \texttt{perceptual score} and subjective evaluation.
Other observation shows the explicit perceptual loss tends to generate high frequency artifacts which is considered as a noise on human perception as shown in Fig.~\ref{fig:subject}.

\section{Concluding Remarks}
\label{section:conclusion}

We proposed an explicit way to utilize machine learning model which trained to produce the perceptual score on generated SR images.s
The experimental results show the proposed approach is able to improve the perceptual score better than the implicit approaches.
We further show the characteristic of implicit and explicit perceptual loss for easier interpretation.
We also demonstrate that the existing perceptual score does not correlate well with human perception using subjective evaluation.
The results open more challenges to create better image quality metrics which can be used explicitly to optimize the SR network.

Future work includes an extension of this work to other SR problems
(i.e., video SR~\cite{DBLP:conf/cvpr/NahTGBHMSL19,DBLP:conf/cvpr/HarisSU19,eccv2020vsr}, time
SR~\cite{DBLP:conf/cvpr/JiangSJ0LK18,eccv2020tsr}, and space-time
SR~\cite{DBLP:conf/cvpr/HarisSU20,DBLP:conf/cvpr/XiangTZ0AX20}).

This work is partly supported by 
JSPS KAKENHI Grant Number 19K12129.

\bibliographystyle{ieee}
\begin{small}
\bibliography{mva2019-psr}

\begin{thebibliography}{10}\itemsep=-1pt

\bibitem{Agustsson_2017_CVPR_Workshops}
E.~Agustsson and R.~Timofte.
\newblock Ntire 2017 challenge on single image super-resolution: Dataset and
  study.
\newblock In {\em The IEEE Conference on Computer Vision and Pattern
  Recognition (CVPR) Workshops}, July 2017.

\bibitem{DBLP:journals/corr/abs-1809-07517}
Y.~Blau, R.~Mechrez, R.~Timofte, T.~Michaeli, and L.~Zelnik{-}Manor.
\newblock 2018 {PIRM} challenge on perceptual image super-resolution.
\newblock {\em arXiv}, abs/1809.07517, 2018.

\bibitem{pirm2018}
Y.~Blau, R.~Mechrez, R.~Timofte, T.~Michaeli, and L.~Zelnik-Manor.
\newblock 2018 pirm challenge on perceptual image super-resolution.
\newblock {\em arXiv preprint arXiv:1809.07517}, 2018.

\bibitem{DBLP:conf/cvpr/BlauM18}
Y.~Blau and T.~Michaeli.
\newblock The perception-distortion tradeoff.
\newblock In {\em CVPR}, 2018.

\bibitem{DBLP:conf/iccv/0002YXD19}
X.~Deng, R.~Yang, M.~Xu, and P.~L. Dragotti.
\newblock Wavelet domain style transfer for an effective perception-distortion
  tradeoff in single image super-resolution.
\newblock In {\em ICCV}, pages 3076--3085. {IEEE}, 2019.

\bibitem{dosovitskiy2016generating}
A.~Dosovitskiy and T.~Brox.
\newblock Generating images with perceptual similarity metrics based on deep
  networks.
\newblock In {\em Advances in Neural Information Processing Systems}, pages
  658--666, 2016.

\bibitem{DBLP:conf/icip/FangLZLG16}
Y.~Fang, J.~Liu, Y.~Zhang, W.~Lin, and Z.~Guo.
\newblock Quality assessment for image super-resolution based on energy change
  and texture variation.
\newblock In {\em ICIP}, 2016.

\bibitem{gatys2016image}
L.~A. Gatys, A.~S. Ecker, and M.~Bethge.
\newblock Image style transfer using convolutional neural networks.
\newblock In {\em Proceedings of the IEEE Conference on Computer Vision and
  Pattern Recognition}, pages 2414--2423, 2016.

\bibitem{goodfellow2014generative}
I.~Goodfellow, J.~Pouget-Abadie, M.~Mirza, B.~Xu, D.~Warde-Farley, S.~Ozair,
  A.~Courville, and Y.~Bengio.
\newblock Generative adversarial nets.
\newblock In {\em Advances in neural information processing systems}, pages
  2672--2680, 2014.

\bibitem{DBLP:conf/iccvw/GuLZXYZYSTDLDLG19}
S.~Gu et~al.
\newblock {AIM} 2019 challenge on image extreme super-resolution: Methods and
  results.
\newblock In {\em ICCV Workshop}, pages 3556--3564. {IEEE}, 2019.

\bibitem{DBLP:conf/cvpr/HarisSU18}
M.~Haris, G.~Shakhnarovich, and N.~Ukita.
\newblock Deep back-projection networks for super-resolution.
\newblock In {\em CVPR}, 2018.

\bibitem{DBLP:journals/corr/abs-1904-05677}
M.~Haris, G.~Shakhnarovich, and N.~Ukita.
\newblock Deep back-projection networks for single image super-resolution.
\newblock {\em arXiv}, abs/1904.05677, 2019.

\bibitem{DBLP:conf/cvpr/HarisSU19}
M.~Haris, G.~Shakhnarovich, and N.~Ukita.
\newblock Recurrent back-projection network for video super-resolution.
\newblock In {\em CVPR}, pages 3897--3906. Computer Vision Foundation / {IEEE},
  2019.

\bibitem{DBLP:conf/cvpr/HarisSU20}
M.~Haris, G.~Shakhnarovich, and N.~Ukita.
\newblock Space-time-aware multi-resolution video enhancement.
\newblock In {\em CVPR}, pages 2856--2865. {IEEE}, 2020.

\bibitem{he2015delving}
K.~He, X.~Zhang, S.~Ren, and J.~Sun.
\newblock Delving deep into rectifiers: Surpassing human-level performance on
  imagenet classification.
\newblock In {\em Proceedings of the IEEE International Conference on Computer
  Vision}, pages 1026--1034, 2015.

\bibitem{Irani:1991:IRI:108693.108696}
M.~Irani and S.~Peleg.
\newblock Improving resolution by image registration.
\newblock {\em CVGIP: Graph. Models Image Process.}, 53(3):231--239, Apr. 1991.

\bibitem{eccv2020vsr}
T.~Isobe, X.~Jia, S.~Gu, S.~Li, S.~Wang, and Q.~Tian.
\newblock Video super-resolution with recurrent structure-detail network.
\newblock In {\em ECCV}, 2020.

\bibitem{DBLP:conf/cvpr/JiangSJ0LK18}
H.~Jiang, D.~Sun, V.~Jampani, M.~Yang, E.~G. Learned{-}Miller, and J.~Kautz.
\newblock Super slomo: High quality estimation of multiple intermediate frames
  for video interpolation.
\newblock In {\em CVPR}, pages 9000--9008. {IEEE} Computer Society, 2018.

\bibitem{johnson2016perceptual}
J.~Johnson, A.~Alahi, and L.~Fei-Fei.
\newblock Perceptual losses for real-time style transfer and super-resolution.
\newblock In {\em European Conference on Computer Vision}, pages 694--711.
  Springer, 2016.

\bibitem{DBLP:conf/cvpr/KangYLD14}
L.~Kang, P.~Ye, Y.~Li, and D.~S. Doermann.
\newblock Convolutional neural networks for no-reference image quality
  assessment.
\newblock In {\em CVPR}, 2014.

\bibitem{DBLP:conf/cvpr/LedigTHCCAATTWS17}
C.~Ledig, L.~Theis, F.~Huszar, J.~Caballero, A.~Cunningham, A.~Acosta, A.~P.
  Aitken, A.~Tejani, J.~Totz, Z.~Wang, and W.~Shi.
\newblock Photo-realistic single image super-resolution using a generative
  adversarial network.
\newblock In {\em CVPR}, 2017.

\bibitem{eccv2020sisr}
A.~Lugmayr, M.~Danelljan, L.~V. Gool, and R.~Timofte.
\newblock Srflow: Learning the super-resolution space with normalizing flow.
\newblock In {\em ECCV}, 2020.

\bibitem{DBLP:conf/icip/Luo04}
H.~Luo.
\newblock A training-based no-reference image quality assessment algorithm.
\newblock In {\em ICIP}, 2004.

\bibitem{DBLP:journals/cviu/MaYY017}
C.~Ma, C.~Yang, X.~Yang, and M.~Yang.
\newblock Learning a no-reference quality metric for single-image
  super-resolution.
\newblock {\em Computer Vision and Image Understanding}, 158:1--16, 2017.

\bibitem{DBLP:journals/tip/MaLZDWZ18}
K.~Ma, W.~Liu, K.~Zhang, Z.~Duanmu, Z.~Wang, and W.~Zuo.
\newblock End-to-end blind image quality assessment using deep neural networks.
\newblock {\em {IEEE} Trans. Image Processing}, 27(3):1202--1213, 2018.

\bibitem{DBLP:journals/spl/MittalSB13}
A.~Mittal, R.~Soundararajan, and A.~C. Bovik.
\newblock Making a "completely blind" image quality analyzer.
\newblock {\em {IEEE} Signal Process. Lett.}, 20(3):209--212, 2013.

\bibitem{DBLP:conf/cvpr/NahTGBHMSL19}
S.~Nah et~al.
\newblock {NTIRE} 2019 challenge on video super-resolution: Methods and
  results.
\newblock In {\em CVPR Workshop}, pages 1985--1995. Computer Vision Foundation
  / {IEEE}, 2019.

\bibitem{DBLP:conf/icip/ReibmanBG06}
A.~R. Reibman, R.~M. Bell, and S.~Gray.
\newblock Quality assessment for super-resolution image enhancement.
\newblock In {\em ICIP}, 2006.

\bibitem{bib:NSS}
D.~L. Ruderman.
\newblock The statistics of natural images.
\newblock {\em Network Computation in Neural Syst.}, 5:517--548, 1994.

\bibitem{DBLP:conf/iccv/SajjadiSH17}
M.~S.~M. Sajjadi, B.~Sch{\"{o}}lkopf, and M.~Hirsch.
\newblock Enhancenet: Single image super-resolution through automated texture
  synthesis.
\newblock In {\em ICCV}, 2017.

\bibitem{simonyan2014very}
K.~Simonyan and A.~Zisserman.
\newblock Very deep convolutional networks for large-scale image recognition.
\newblock {\em ICLR}, 2015.

\bibitem{DBLP:conf/cvpr/TangJK14}
H.~Tang, N.~Joshi, and A.~Kapoor.
\newblock Blind image quality assessment using semi-supervised rectifier
  networks.
\newblock In {\em CVPR}, 2014.

\bibitem{DBLP:conf/cvpr/TimofteGWG18}
R.~Timofte, S.~Gu, J.~Wu, and L.~V. Gool.
\newblock {NTIRE} 2018 challenge on single image super-resolution: Methods and
  results.
\newblock In {\em {CVPR} Workshop}, 2018.

\bibitem{DBLP:journals/tip/WangBSS04}
Z.~Wang, A.~C. Bovik, H.~R. Sheikh, and E.~P. Simoncelli.
\newblock Image quality assessment: from error visibility to structural
  similarity.
\newblock {\em {IEEE} Trans. Image Processing}, 13(4):600--612, 2004.

\bibitem{DBLP:conf/cvpr/XiangTZ0AX20}
X.~Xiang, Y.~Tian, Y.~Zhang, Y.~Fu, J.~P. Allebach, and C.~Xu.
\newblock Zooming slow-mo: Fast and accurate one-stage space-time video
  super-resolution.
\newblock In {\em CVPR}, pages 3367--3376. {IEEE}, 2020.

\bibitem{DBLP:conf/icip/YeganehRW12}
H.~Yeganeh, M.~Rostami, and Z.~Wang.
\newblock Objective quality assessment for image super-resolution: {A} natural
  scene statistics approach.
\newblock In {\em ICIP}, 2012.

\bibitem{DBLP:conf/cvpr/ZhangGTSDZYGJYK20}
K.~Zhang et~al.
\newblock {NTIRE} 2020 challenge on perceptual extreme super-resolution:
  Methods and results.
\newblock In {\em CVPR Workshop}, pages 2045--2057. {IEEE}, 2020.

\bibitem{eccv2020tsr}
L.~P. Zuckerman, E.~Naor, G.~Pisha, S.~Bagon, and M.~Irani.
\newblock Across scales \& across dimensions: Temporal super-resolution using
  deep internal learning.
\newblock In {\em ECCV}, 2020.

\end{thebibliography}
\end{small}

\end{document}